\documentclass[twocolumn,showpacs,preprintnumbers,aps,prd,a4paper,
superscriptaddress,nofootinbib,tightenlines,floats,floatfix]{revtex4}
\usepackage{graphicx}

\usepackage{latexsym}

\usepackage{amsmath,amssymb}        

\usepackage[draft=false]{hyperref}

\usepackage{dcolumn}
\usepackage{bm}
\usepackage{color}

\def\lsim{\mathrel{\mathop
  {\hbox{\lower0.5ex\hbox{$\sim$}\kern-0.8em\lower-0.7ex\hbox{$<$}}}}}
\def\gsim{\mathrel{\mathop
  {\hbox{\lower0.5ex\hbox{$\sim$}\kern-0.8em\lower-0.7ex\hbox{$>$}}}}}

\begin{document}

\newcommand{\1}{$\spadesuit$}
\newcommand{\half}{{1\over2}}
\newcommand{\nad}{n_{\rm ad}}
\newcommand{\niso}{n_{\rm iso}}
\newcommand{\ncor}{n_{\rm cor}}
\newcommand{\fiso}{f_{\rm iso}}
\newcommand{\ii}{\'{\'i}}
\newcommand{\bk}{{\bf k}}
\newcommand{\Ocdm}{\Omega_{\rm cdm}}
\newcommand{\ocdm}{\omega_{\rm cdm}}
\newcommand{\OM}{\Omega_{\rm M}}
\newcommand{\OB}{\Omega_{\rm B}}
\newcommand{\oB}{\omega_{\rm B}}
\newcommand{\OX}{\Omega_{\rm X}}
\newcommand{\cltt}{C_l^{\rm TT}}
\newcommand{\clte}{C_l^{\rm TE}}
\newcommand{\calR}{{\cal R}}
\newcommand{\calS}{{\cal S}}
\newcommand{\Rrad}{{\cal R}_{\rm rad}}
\newcommand{\Srad}{{\cal S}_{\rm rad}}
\newcommand{\calPR}{{\cal P}_{\cal R}}
\newcommand{\calPS}{{\cal P}_{\cal S}}

\input epsf

\preprint{astro-ph/0501477}
\title{Bayesian model selection and isocurvature perturbations}
\author{Mar\'\i a Beltr\'an}
\affiliation{Departamento de F\'\i sica Te\'orica \ C-XI, Universidad
Aut\'onoma de Madrid, Cantoblanco, 28049 Madrid, Spain}
\affiliation{Astronomy Centre, University of Sussex, 
Brighton BN1 9QH, United Kingdom}
\author{Juan Garc{\'\i}a-Bellido}
\affiliation{Departamento de F\'\i sica Te\'orica \ C-XI, Universidad
Aut\'onoma de Madrid, Cantoblanco, 28049 Madrid, Spain}
\author{Julien Lesgourgues}
\affiliation{LAPTH, F-74941 Annecy-le-Vieux Cedex, France
  and INFN Padova, Via Marzolo 8, I-35131 Padova, Italy}
\author{Andrew R. Liddle}
\affiliation{Astronomy Centre, University of Sussex, Brighton BN1 9QH, United 
Kingdom}
\author{An\u{z}e Slosar}
\affiliation{Faculty of Mathematics and Physics, University of Ljubljana, 
Slovenia}
\date{\today}
\pacs{98.80.Cq \hfill IFT-UAM/CSIC-05-02, LAPTH-1085/05, DFPD/05/A05, 
astro-ph/0501477}
\begin{abstract}
Present cosmological data are well explained assuming purely adiabatic
perturbations, but an admixture of isocurvature perturbations is also
permitted.  We use a Bayesian framework to compare the performance of
cosmological models including isocurvature modes with the purely
adiabatic case; this framework automatically and consistently penalizes
models which use more parameters to fit the data. We compute the
Bayesian evidence for fits to a dataset comprised of WMAP and other microwave
anisotropy data, the galaxy power spectrum from 2dFGRS and SDSS, and Type Ia 
supernovae luminosity distances. We
find that Bayesian model selection favours the purely adiabatic models,
but so far only at low significance.
\end{abstract}

\maketitle

\section{Introduction}

Following recent developments in observational cosmology, particularly 
observations by the Wilkinson Microwave Anisotropy Probe (WMAP) \cite{WMAP}, 
there exist
compelling reasons to talk about a Standard Cosmological Model based on
the $\Lambda$CDM paradigm seeded with purely adiabatic perturbations. In
addition, there have been many attempts to analyze more general models
featuring additional physics, either to constrain such processes or in
the hope of discovering some trace effects in the data. A case of
particular interest is the possible addition of an admixture of
isocurvature perturbations to the adiabatic ones
\cite{Efstathiou:1986,Bucher:1999re} which has been studied in the
post-WMAP era by many authors
\cite{wmapp,Gordon:2002gv,Valiviita:2003ty,Crotty:2003aa,Gordon:2003hw,
Parkinson,BGLR,KMV}.

The bulk of investigations so far have as a starting point chosen a particular 
set of parameters to define the cosmological model under discussion, and then 
attempted to constrain those parameters using observations, a process known as 
{\em parameter fitting}. Based on such analyses, many parameters are determined 
to a high degree of accuracy. Much less attention has been directed at the 
higher-level inference problem of allowing the data to decide the set of 
parameters to be used, known as {\em model comparison} or {\em model selection} 
\cite{jeff,mackay}, 
although such techniques have been widely deployed outside of astrophysics. 
Recently, one of us applied two model selection statistics, known as the Akaike 
and Bayesian Information Criteria, to some simple cosmological models 
\cite{Liddle:2004nh}, and 
showed that the simplest model considered was the one favoured by the data. 
These criteria have recently been applied to models with isocurvature 
perturbations by Parkinson et al.~\cite{Parkinson}, who concluded that the 
purely adiabatic 
model was favoured.

Those statistics are not however full implementations of Bayesian inference,
which appears to be the most appropriate framework for interpretting
cosmological data.  The correct model selection tool to use in that context is
the {\em Bayesian evidence} \cite{jeff,mackay}, which is the probability of the
model in light of the data (i.e~the average likelihood over the prior
distribution). It has been deployed in cosmological contexts by several
authors \cite{evidence}, and the ratio of evidences between two models is also 
known as the Bayes Factor \cite{KR95}.\footnote{The difference in Bayesian 
Information Criterion can be used as a
crude approximation to $\ln(\mbox{Bayes Factor})$, but the existence of 
parameter degeneracies in
cosmological data fitting are likely to violate the conditions for the validity
of the approximation.}   The Bayesian evidence can be combined with prior
probabilities for different models if desired, but even if the prior
probabilities are assumed equal, the evidence still automatically encodes a
preference for simpler models, implementing Occam's razor in a quantitative
manner.

Whenever one aims to decide whether or not a particular parameter $p$ should be 
fixed
(for example at $p=0$), one should use model selection techniques.
If one carries out only a parameter-fitting exercise and then examines the 
likelihood level at which $p=0$ is excluded, such a comparison fails to account 
for the model dimensionality being reduced by one at the point $p=0$, and hence 
draws conclusions inconsistent with Bayesian inference. This typically 
overestimates
the significance at which the parameter $p$ is needed. 
An example is spectral index running, which parameter fitting favours at a
modest (albeit unconvincing) confidence level \cite{WMAP}, but which is 
disfavoured by model selection statistics \cite{Liddle:2004nh}.

In this paper we use the Bayesian evidence to compare isocurvature and adiabatic 
models in light of current data. We will closely follow the notation of 
Beltr\'an et al.~\cite{BGLR}, who recently carried out a parameter-fitting 
analysis of 
isocurvature models, and we use the same datasets. We follow the notation of 
that paper and provide only a brief summary in this article.

\section{Bayesian evidence}

\subsection{Theoretical basis}

The Bayesian evidence is the average likelihood of a model over its prior 
parameter space, namely
\begin{equation}
\label{e:evidencedef}
E = \int {\cal L}(\mathbf{\theta})\; {\rm pr}(\mathbf{\theta})\; 
d\mathbf{\theta} \,,
\end{equation}
where $\mathbf{\theta}$ is the parameter vector defining the model, ${\rm 
pr}(\mathbf{\theta})$ the normalized priors on those parameters (typically taken 
to be top-hat distributions over some range), and ${\cal L}(\mathbf{\theta})$ is 
the likelihood. In essence, it asks the question: `If I consider the possible 
model parameters I was allowing before I knew about this data, on average how 
well did they fit the data?'. Generally speaking, models with fewer 
parameters tend to be more predictive, and provided that for some parameter 
choices they fit the data well, then the average likelihood can be expected to 
be higher. On the other hand, a simple model which cannot fit the data for any 
parameter choices will not generate a good likelihood. The Bayesian evidence 
therefore sets up the desired tension between model simplicity and ability to 
explain the data.

Models are ranked in order of their Bayesian evidence, usually using
its logarithm. The overall normalization is irrelevant. As the
evidence is the (unnormalized) probability of the model, if two models
are being compared, the odds of the one with the lower evidence is
$1/(1+\exp(\Delta\ln E))$. What constitutes a significant difference
is to some extent a matter of personal taste, but a useful guide is given by 
Jeffreys \cite{jeff} who rates $\Delta \ln E < 1$ as `not worth more than a bare 
mention', $1<\Delta \ln E < 2.5$ as `substantial', $2.5<\Delta \ln E < 
5$ `strong' to `very strong' and $5<\Delta \ln E$ as `decisive', in each case 
the decision being against the model with the smaller evidence. Note that a 
difference $\Delta \ln
E$ of 2.5 corresponds to odds of 1 in about 13, and
$\Delta \ln
E$ of 5 to odds of 1 in $150$.

A significant, but unavoidable, disadvantage of the use of the evidence is that
it depends on the prior ranges chosen for the parameters.  For
instance, if one doubles the range of one parameter by allowing it to
vary in a region where the likelihood is negligibly small, then the
evidence will half.  Indeed, one can make any model disfavoured
simply by extending its prior range indefinitely in a direction where
there is no hope of fitting the data. From a Bayesian point of view
this is unsurprising; of course your belief in a model should be
influenced by what you thought of it before the data came along, and
the Bayesian analysis has the virtue of forcing you to make your
assumptions explicit.

However, the prior width is not as crucial as
one might na\"{\i}vely expect. The main reason is that the likelihood is
typically falling off exponentially away from the best fit, while the parameter
volume is growing only as a polynomial function. For example,
consider a one-dimensional toy-model for which the likelihood is given
by
\begin{equation}
{\cal L} (x) = {\cal L}_0 \exp \left( -\frac{(x-\mu)^2}{2}\right)\,,
\end{equation}
and consider two models: model A is $x=0$ and model B is $x\neq 0$ with a
top-hat prior $0<x<a$. In the case $\mu=1$, a conventional 1-$\sigma$  
non-detection,
the evidence would be unable to strongly distinguish between the
models ($|\Delta \ln E| < 2.5$) for up to $a \sim 50$. In the case $\mu=5$,
a conventional 5-$\sigma$ detection, the evidence would prefer model
B for all $a \lsim 5 \times 10^4$. In other words, for reasonable prior
ranges the evidence will robustly pick up the correct model. Its main
advantage is that it is a quantitative measure with clear
interpretation within Bayesian statistics, and can be applied in cases
where the usual frequentist arguments do not provide us with definite answers.
Typically, Bayesian analysis contradicts the frequentist results whenever the 
latter accepts a parameter in light of a marginally better $\chi^2$ value. If 
this 
improvement 
is not significant, the increase of the volume of the parameter space
will penalize the addition of the new parameter and thus 
decrease the evidence of the extended model.

Generally the evidence is not reparametrization invariant, in the
sense that the choice of a flat prior in one parametrization will
probably not correspond to a flat prior under another
parametrization. The choice of parametrization is a matter of personal 
preference, though obviously truly robust model
selection results should be preserved under reasonable changes in
parametrization. In the case of isocurvature perturbations there are
different, equally plausible, choices of parametrization, in particular geared 
to dealing
with the problem of the cross-correlation angle becoming unconstrained
as the isocurvature mode amplitude becomes small \cite{BGLR,KMV}. 
For illustration we will compare the results obtained under two 
different parametrization choices.

\subsection{Numerical implementation}

The evidence for a given model can be computed by a Markov Chain Monte Carlo 
method. However it cannot be directly calculated from chains used in parameter 
estimation (for instance from the program {\tt CosmoMC} \cite{cosmomc}), because 
those chains are sampled from the posterior distribution, which is peaked around 
the maximum likelihood, and do not carry the necessary information on the 
likelihood far from the maximum. Equally, one cannot simply sample from the 
prior distribution, because the dominant contribution from the high-likelihood 
regions will not be properly sampled. Consequently, a hybrid technique is 
required, a useful method being thermodynamic integration 
\cite{thermo,Hobson2002}.

Thermodynamic integration alters the sampling of a Markov chain by introducing a 
parameter $\lambda$, thought of as an inverse temperature, with the acceptance 
rate governed by the likelihood raised to the power $\lambda$. As $\lambda$ is 
varied from zero to one, this interpolates between sampling from the prior and 
the posterior distributions. Defining 
\begin{equation}
E(\lambda) = \int {\cal L}^\lambda (\mathbf{\theta})\; {\rm 
pr}(\mathbf{\theta})\; d\mathbf{\theta} \,,
\end{equation}
it can be shown that 
\begin{equation}
\ln E = \ln \frac{E(1)}{E(0)}  =   \int_0^1 \frac{d\ln E}{d\lambda} \, 
d\lambda  = \int_0^1 \langle \ln {\cal L} \rangle_\lambda \, d\lambda \,,
\label{eq:ev}
\end{equation}
where
\begin{equation}
\label{e:integrand}
\langle \ln {\cal L} \rangle_\lambda \equiv \frac{\int \ln {\cal L} \; {\cal 
L}^\lambda \, {\rm pr}(\mathbf{\theta}) \, d \mathbf{\theta}}{\int {\cal 
L}^\lambda \, {\rm pr}(\mathbf{\theta}) \, d \mathbf{\theta}}
\end{equation}
is the average of $\ln {\cal L}$ over the distribution at temperature
$1/\lambda$. That the priors in Eq.~(\ref{e:evidencedef}) must be normalized 
implies that $E(0)$ equals one, though the prior normalization anyway cancels 
out in the integrand Eq.~(\ref{e:integrand}).

Previous work in cosmology has typically evaluated the evidence during the 
burn-in
phase of a chain to be used for parameter estimation. In this process,
the temperature is slowly cooled from $\lambda = 0$ to $\lambda=1$ to
facilitate the relaxation of the chain into its stationary
distribution and those chain elements are used for evidence
computation; they are then discarded and the remaining elements, all
sampled at $\lambda = 1$, are used for parameter estimation. This
method is ideal for complex inference problems with dimensionality
$d\gg 1$ and multimodal likelihood distributions, where a slow burn-in
phase is necessary to explore the posterior in an unbiased manner and
thus the evidence calculation comes `for free'.  However,  in a typical
cosmological problem the likelihood surface is considerably simpler, arguably 
unimodal, and
the number of samples required for a reliable burn-in is much smaller
than the number of samples needed for an accurate evidence
estimation. Therefore, we choose a different approach in which we heat
the chain, using the endpoint of a parameter estimation run as the starting 
point. Since the volume of parameter space is larger at
higher temperatures it should be much easier
to ensure that the chain is stationary at each temperature step during
heating rather than cooling. We implemented two
different heating schedules:
\begin{itemize}
\item \textit{Continuous temperature change.}  We let the
inverse sampling temperature change continuously at each step as
\begin{equation}
  \lambda (n) = (1-\xi)^n,
\end{equation}
where $n$ is the step number. The single sample taken at that
temperature can be viewed as an unbiased (although noisy) estimate of
$\langle \ln {\cal L} \rangle_\lambda$. This continuous approach obviates 
the problem of deciding the number of steps per
position, transferring it to the step size. When the algorithm decides
to stop, the integral is closed to $\lambda=0$ in the last step. The
stopping criterion is that the closure of the integral by the last step
would change $\ln E$ by less than
a certain threshold, $\epsilon_{\rm stop}$, even for the most extreme
likelihood encountered. The choices of $\xi$ and $\epsilon_{\rm stop}$
determine the accuracy and speed of the evidence calculator, and optimum
values must be determined empirically.  After trying various
possibilities we settled for $\xi=5\times 10^{-5}$ and $\epsilon_{\rm
stop} = 0.001$. We have tested that decreasing either $\xi$ or
$\epsilon_{\rm stop}$ further does not affect our results.

\item \textit{Stepwise temperature change.} The integrand of Eq.~(\ref{eq:ev}) 
is first estimated at $\lambda=1$ and $0$, then at intermediate temperatures
given by
\begin{equation}
    \lambda_n = \frac{1}{q^n}\,,
\end{equation}
($q$ is typically 1.5 -- 2 and $n$ an increasing integer). The 
thermodynamic integral is
calculated by the trapezoid rule after each additional point is
added. The points are added until the integral converges to a user-specified 
stopping accuracy $\epsilon_{\rm stop}$. At each
temperature the integral is calculated by making a short burn-in at
that temperature (typically 400 samples, since the chain must
already be roughly burned in from the previous step) and then
calculating $\langle \ln {\cal L} \rangle_\lambda$ from a further
number (typically 1000) of \emph{accepted} samples. This approach has the 
disadvantage that extra samples are needed for burn-in at
each temperature and that there might be systematics associated with
stepwise temperature change. However, it is less sensitive to
the quality of covariance matrix as a poorer covariance matrix
simply results in more samples being taken to get enough accepted
samples (note that we cannot do the same for the continuous
scheme without biasing the result, unless one is
willing to burn-in at \emph{each} `continuous' temperature
change step).
\end{itemize}
Additionally, we modify the proposal function so that its
width scales with $\lambda^{-1}$ (up to a certain width), which
ensures that at high temperatures the chain is sampling randomly from
the prior, rather than random-walking with the step-size corresponding
to the $\lambda=1$ posterior.

These two methods have been extensively tested to give results that
are consistent and accurate to within a unit of $\ln E$ for a single
run. The final numbers for all models were calculated using the
continuous temperature change method.  Additionally we have performed a
comparison with an analytic approximation to the posterior and got results that 
are also consistent to better than one unit of
$\ln E$ in the adiabatic case, though slightly worse in the isocurvature case.

In all cases we find that the number of samples required to accurately
estimate the evidence and avoid systematics associated with covariance
matrices, proposal widths and similar is unexpectedly large; an order of
magnitude larger than what is required for a simple parameter
estimation. This makes the computation a challenging task as it is limited by 
the speed of the likelihood evaluations which require generation of the model 
power spectra.
This also suggests that the uncertainties on evidence
values already found in the literature may be underestimated, though we
note that the high quality of the WMAP data makes this task
considerably more difficult than it was in the pre-WMAP era. Further 
investigation into evidence
estimation methods is clearly warranted and will be a focus of a 
forthcoming paper.

\section{Evidence for isocurvature models}

Our principal aim is to compare the evidence of isocurvature models
with purely adiabatic ones. We will follow the notation of Beltr\'an
et al.~\cite{BGLR}. In general there are four types of isocurvature
modes \cite{Bucher:1999re} --- cold dark matter isocurvature (CDI),
baryon isocurvature (BI), neutrino isocurvature density (NID), and
neutrino isocurvature velocity (NIV) --- but the first two are
observationally indistinguishable \cite{Gordon:2002gv} so we ignore the baryon
isocurvature case. These modes can exist in any combination, and with
correlations both amongst themselves and with the adiabatic modes. We
will only allow a single type of isocurvature mode in any model,
though we will allow a general spectral index both for the
isocurvature modes and for their correlation with the adiabatic ones.

\begin{table}[t]
\begin{center}
\begin{tabular}{c c c}
\hline
Parameter         &\hspace{3mm}Prior Range\hspace{3mm}    
&\hspace{3mm}Model\hspace{3mm}\\
\hline
$\omega_{\rm b}$            &(0.018,0.032)                &AD-HZ,AD-$n_{{\rm 
s}}$,ISO       \\
$\omega_{\rm dm}$         &(0.04,0.16)                  &AD-HZ,AD-$n_{{\rm 
s}}$,ISO       \\
$\theta$              &(0.98,1.10)                  &AD-HZ,AD-$n_{{\rm s}}$,ISO  
     \\
$\tau$                &(0,0.5)                      &AD-HZ,AD-$n_{{\rm s}}$,ISO  
     \\
$\ln[10^{10}{\Rrad}]$ &(2.6,4.2)                    &AD-HZ,AD-$n_{{\rm s}}$,ISO  
     \\
$n_{\rm s}$                 & (0.8,1.2)                   &AD-$n_{{\rm s}}$,ISO  
     \\
$n_{\rm iso}$             & (0,3)                       &ISO       \\
$\delta_{\rm cor}$        & ($-$0.14,0.4)                 &ISO       \\
$\sqrt{\alpha}$              & ($-$1,1)                      &ISO       \\
$\beta$               & ($-$1,1)                      &ISO       \\
\hline
\end{tabular}
\end{center}
\caption{\label{params} The parameters used in the models. The sound horizon 
$\theta$ was used in place of the Hubble parameter. For the AD-HZ model $n_{\rm 
s}$ was fixed to $1$ and
$n_{\rm iso}$, $\delta_{{\rm cor}}$, $\alpha$ and $\beta$ were fixed to $0$. In 
the AD-$n_{{\rm s}}$ model, $n_{{\rm s}}$ also varies. Every isocurvature model 
holds the same priors
for the whole set of parameters.}
\end{table}

The flat prior ranges for all parameters are given in Table~\ref{params}. 
We consider two adiabatic models. AD-HZ is the simplest model giving a 
good fit to the data, with a Harrison--Zel'dovich spectrum and five 
variable parameters. We also computed the evidence for 
an extended adiabatic model AD-$n_{{\rm s}}$ in which we let $n_{s}$ vary.

For each isocurvature model there are four
extra parameters.
As in Ref.~\cite{BGLR} we parametrize the contribution to the temperature
and polarization angular power spectra from the adiabatic,
isocurvature and correlation amplitudes at the pivot scale ($k_0=0.05$
Mpc$^{-1}$) by $\alpha$ and $\beta$ so that:
\begin{equation}
C_l = (1-\alpha)\,C_l^{\rm ad} + \alpha\,C_l^{\rm iso} + 2\beta\,
\sqrt{\alpha(1-\alpha)}\,C_l^{\rm cor}\,.
\end{equation}
The parameter $\delta_{{\rm cor}}$ is related to the spectral tilt of the
correlation mode, $n_{\rm cor}$, and its boundaries are fixed by the
pivot scale and the $k_{{\rm min}}=4 \times 10^{-5}$ Mpc$^{-1}$ and
$k_{{\rm max}}=0.5$ Mpc$^{-1}$ scales used for the analysis. It is defined as
\begin{equation}
\delta_{\rm cor} \equiv n_{\rm cor} / \ln |\beta|^{-1} \,.
\end{equation}
Thus the priors on the first seven parameters are theoretically
motivated, whereas the priors on the last three are automatically set
by the model.
Throughout the analysis the equation of state parameter of the
dark energy was set to $-1$.

We have used 
the following datasets: cosmic microwave anisotropy data from the WMAP satellite 
including temperature--polarization cross-correlation
\cite{WMAP}, VSA \cite{VSA}, CBI \cite{CBI} and ACBAR \cite{ACBAR}, matter power 
spectrum data from the two-degree field galaxy redshift 
survey (2dFGRS) power spectrum
\cite{2dFGRS} and from the Sloan Digital Sky Survey \cite{SDSS}, and the 
supernovae apparent magnitude--redshift relation \cite{Riess2004}.

\begin{table}[t]
\begin{center}
\begin{tabular}{c c r}
\hline
Model              &\hspace{5mm}      &\hspace{3mm}ln(Evidence)\hspace{5mm}\\
\hline
AD-HZ                 &\hspace{5mm}       & 0.0  $\pm$ 0.1\\
AD-$n_{{\rm  s}}$      &\hspace{5mm}       & 0.0  $\pm$ 0.1\\
CDI                &\hspace{5mm}       & $-$1.0  $\pm$ 0.2\\
NID                &\hspace{5mm}       & $-$1.0  $\pm$ 0.2\\
NIV                &\hspace{5mm}       & $-$1.0  $\pm$ 0.3\\
\hline
\end{tabular}
\end{center}
\caption{\label{evid} Evidences for the four different models studied,
normalized to the AD-HZ evidence. The absolute value for that model was
$\ln E = -854.1$.}
\end{table}

We ran 32 independent computations of the evidence for each model.  In all of
them the stopping criterion was satisfied after about $2.5\times 10^5$ steps, so
the total number of likelihood evaluations was approximately $10^7$ per model.
The results, given as the logarithm of the evidence, are described in
Table~\ref{evid}.  We have expressed all the calculated evidence values relative
to the AD-HZ model, as the absolute value is just a particular of the likelihood
code.  We see from the table that the evidences are calculated to sufficient
accuracy to draw conclusions, but that the comparison is rather inconclusive.
Firstly, the two adiabatic models happen to produce the same evidence; as a 
further consistency check, we also looked at an adiabatic model with the
prior range on $n_{\rm s}$ doubled, and found that $\ln E$ fell by 0.4, to be 
compared with the expected drop of $\ln 2$ that would appear if the likelihood
were insignificant throughout the extended range.  Secondly, by
coincidence all three isocurvature models have the same evidence, with $\Delta
\ln E$ being 1.0 relative to AD-HZ in each case.  According to the Jeffreys'
scale this is just at the edge of being worthy of attention.

As mentioned in Section~II, these results are not reparametrization
invariant, since changing the basis of parameters typically leads to a
different choice of priors. Various parameterizations have been used in
the literature. For instance, a change of pivot scale leads to an
$(n_\mathrm{s}-n_\mathrm{iso})$--dependent rescaling of $\alpha$, and to
an $n_\mathrm{cor}$--dependent rescaling of $\beta$. Even if the pivot
scale is fixed, various definitions of the amplitude parameters can be
introduced.  The normalization of the isocurvature mode can be
parametrized by the ratio of isocurvature to adiabatic primordial
fluctuations $f_{\rm iso} \in [0,\infty]$ \cite{wmapp} instead of the
fraction of isocurvature contribution to the total primordial spectrum
$\alpha \in [0,1]$~\cite{Crotty:2003aa}.  In this work, as in
Ref.~\cite{BGLR}, we chose to vary $\sqrt{\alpha} \in [-1,1]$ in order
to avoid dealing with boundary effects and to have a posterior
distribution falling down to zero on the two ends of the prior range. We
could nevertheless instead have chosen a flat prior for $\alpha$.
Similarly, the cross-correlation amplitude can be parametrized either by
the correlation angle $\beta \in [-1,1]$, as in Refs.~\cite{wmapp,BGLR},
or by the amplitude of the cross-correlation power spectrum $2\beta
\sqrt{\alpha(1-\alpha)}$~\cite{Valiviita:2003ty}. The advantage of the
latter is that the total power spectrum depends linearly on it, and so
it is well constrained by the data, while starting from a flat prior on
$\beta$ we can get a flat posterior distribution if the preferred model
is purely adiabatic, so that the value of $\beta$ does not matter (this
point is discussed in detail in Ref.~\cite{KMV} where a third choice is
also introduced). Finally, we defined the parameter
$\delta_\mathrm{cor}$ in order to deal with a simple top-hat prior, but
we could decide to use instead to impose a flat $\beta$--dependent prior
directly on $n_\mathrm{cor}$.

To get a hint of the effect of reparametrization, we recomputed
the evidences using a second parameter basis: instead of
($\sqrt{\alpha}$, $\beta$) we vary ($\alpha$, $2\beta
\sqrt{\alpha(1-\alpha)}$) with a flat prior inside the two-dimensional
ellipse in which these parameters are defined, and instead of
$\delta_\mathrm{cor}$ we vary $n_\mathrm{cor}$ within the range $[-0.14
\ln(|\beta|^{-1}), 0.4 \ln(|\beta|^{-1})]$. Since the prior on
$n_\mathrm{cor}$ is too loose when $\beta$ is close to zero, we imposed
the additional prior over $n_\mathrm{cor} \in [-1,1]$.

\begin{table}[t]
\begin{center}
\begin{tabular}{c c r}
\hline
Model    &\hspace{5mm}      &\hspace{3mm}ln(Evidence)\hspace{5mm}\\
\hline
AD-HZ      &\hspace{5mm}       &  0.0  $\pm$ 0.1\\
CDI     &\hspace{5mm}       & $-$1.0  $\pm$ 0.2\\
NID     &\hspace{5mm}       & $-$2.0  $\pm$ 0.2\\
NIV     &\hspace{5mm}       & $-$2.3  $\pm$ 0.2\\
\hline
\end{tabular}
\end{center}
\caption{\label{evid_newParam} Evidences for the four models using the second 
parametrization, again normalized to the AD-HZ evidence.}
\end{table}

The results are quoted in Table~\ref{evid_newParam}, and show differences
from the ones that use the original parametrization. Even though the
difference is still not big enough to exclude any isocurvature model, we
conclude that, as mentioned in Section II, parametrization does matter
for the evidence calculation.

\section{Conclusions}

We have carefully calculated the evidence for two adiabatic models and
three physically-distinguishable isocurvature models using recent cosmic
microwave background, supernovae and large-scale structure data.  We find very
similar evidences for all the models. For the first parametrization
used, the odds of the isocurvature models compared to the adiabatic ones
are 1 in about 4. Using a second parametrization of the isocurvature
parameters we find the odds for the neutrino cases drop to 1 in 10.  Therefore, 
we conclude
that present data are unable to offer a clear verdict for or against the
inclusion of isocurvature degrees of freedom.  This conclusion is
similar to that found by Parkinson et al.~\cite{Parkinson} using the
information criteria.  Although the extra parameters introduce extra
complexity, these models are still able to satisfactorily fit the
present data for a wide range of their parameters and thus the evidence
quantifies the common sense that one should allow these models to be
considered.  We also showed the relevance of the parametrization for
evidence computation.

While the present comparison is inconclusive, a key question for future
data will be to select between the adiabatic and isocurvature
paradigms. Parameter estimation analyses cannot do this, as even if the
adiabatic model is correct they can only impose limits on the
isocurvature parameters.  The Bayesian model selection approach we have
described is the ideal tool to carry out such a selection.

\begin{acknowledgments}
M.B. was supported at Sussex by a Marie Curie Fellowship of
the European Community programme HUMAN POTENTIAL under contract
HPMT-CT-2000-00096, J.L. in Padova by INFN, and A.R.L. by PPARC.  A.S. 
acknowledges useful discussions with Phil Marshall, Michael
Hobson, and Uro\v{s} Seljak, and support from the Slovene Ministry of
Science, Education and Sport.   This work was supported in
part by a CICYT project FPA2003-04597, and by a Spanish--French
collaborative grant between CICYT and IN2P3. We acknowledge the use of the
COSMOS cluster for our computations with the CosmoMC code, and thank
the sponsors of this UK-CCC facility, supported by HEFCE and
PPARC. This research was conducted in cooperation with GSI/Intel
utilizing the Altix 3700 supercomputer.
\end{acknowledgments}

\end{document}